# Demonstration of a solid deuterium source of ultra-cold neutrons

A. Saunders[*], J. M. Anaya[*], T. J. Bowles[*], B. W. Filippone[†], P. Geltenbort[††], R. E. Hill[*], M. Hino[‡], S. Hoedl[¶], G. E. Hogan[*], T. M. Ito[†], K. W. Jones[*], T. Kawai[‡1], K. Kirch[*2], S. K. Lamoreaux[*], C.-Y. Liu[¶3], M. Makela[**], L. J. Marek[*], J. W. Martin[†], C. L. Morris[*], R. N. Mortensen[*], A. Pichlmaier[*2], S. J. Seestrom[*], A. Serebrov[||], D. Smith[¶4], W. Teasdale[*], B. Tipton[†], R. B. Vogelaar[**], A. R. Young[§], and J. Yuan[†]

[*] *Los Alamos National Laboratory, Los Alamos, NM 87545, USA.*

[†] *Kellogg Radiation Laboratory, California Institute of Technology, Pasadena, Ca 91125, USA.*

[††] *Institut Laue-Langevin, BP 156, F-38042 Grenoble Cedex 9, France.*

[‡] *Research Reactor Institute, Kyoto University, Kumatori, Osaka 590-0494, Japan.*

[§] *North Carolina State University, Raleigh, NC 27695, USA.*

[¶] *Princeton University, Princeton, NJ 08544, USA.*

[**] *Virginia Polytechnical Institute and State University, Blacksburg, Va 24061, USA.*

[||] *St.-Petersburg Nuclear Physics Institute, Russian Academy of Sciences, 188350 Gatchina, Leningrad District, Russia.*

---

[1] Present address: 11-11, 2-chome, Yayoi-cho, Izumi-shi, Osaka 594-0061, Japan.

[2] Present address: Paul Scherrer Institute, CH-5232 Villigen PSI, Switzerland.

[3] Present address: Los Alamos National Lab, Los Alamos, NM 87545, USA.

[4] Present address: Stanford Linear Accelerator Center, Menlo Park, CA 94025, USA.

**Ultra-cold neutrons (UCN), neutrons with energies low enough to be confined by the Fermi potential in material bottles, are playing an increasing role in measurements of fundamental properties of the neutron. The ability to manipulate UCN with material guides and bottles, magnetic fields, and gravity can lead to experiments with lower systematic errors than have been obtained in experiments with cold neutron beams. The UCN densities provided by existing reactor sources limit these experiments. The promise of much higher densities from solid deuterium sources has led to proposed facilities coupled to both reactor and spallation neutron sources. In this paper we report on the performance of a prototype spallation neutron-driven solid deuterium source. This source produced bottled UCN densities of 145 ±7 UCN/cm$^3$, about three times greater than the largest bottled UCN densities previously reported. These results indicate that a production UCN source with substantially higher densities should be possible.**

Free neutrons with velocities below about 8 m/s can be confined within material bottles[1,2,3] and stored for times comparable to the free neutron lifetime. Such neutrons, called Ultra-Cold Neutrons (UCN), have been used to greatly improve measurements of the neutron lifetime[4,5,6] and searches for the neutron electric dipole moment[7,8]. In addition, gravitationally induced quantum states have recently been observed for the first time using UCN[9]. The limitation in all of these measurements is the maximum attainable density of UCN. Existing sources of ultra-cold neutrons (UCN) rely on extracting very low energy neutrons from the Maxwellian tail of the thermal energy distribution in nuclear reactor-driven moderators. In some cases, receding turbine blades, gravity, or both are used to further slow the neutrons. The highest bottled density reported in the literature, obtained at the Institut Laue-Langevin (ILL) reactor in Grenoble, was about 50 UCN/cm³.[10,11] (A density of 87 storable UCN/cm³ upstream of the bottle was also reported.) Golub and Pendlebury[12] suggested that a superfluid liquid helium moderator, in which the production rate of UCN caused by down-scattering in energy is larger than the combined nuclear-absorption (essentially zero for helium-4) and up-scatter rates in the material, could provide higher UCN densities. Golub and Boning[13,14] proposed using thin films of solid deuterium (SD2) as such a superthermal source.

Pokotilovski[15] suggested more effective use could be made of a source with finite absorption by using a pulsed reactor to drive the solid deuterium and by opening and closing a shutter between the source and a storage bottle to reduce the contact time of the UCN with the SD2 to reduce the effect of absorption on the UCN lifetime. Serebrov et al.[16,17] suggested the use of spallation-produced neutrons, which can be produced at higher densities for a given heat load than reactor neutrons, with such a shuttered source.

We have built and tested a prototype spallation neutron-driven SD2 UCN source, shown schematically in Fig. 1. Neutrons were produced by hitting a tungsten spallation target with 800 MeV protons. The neutrons were reflected by a beryllium box held at near liquid nitrogen temperature, then moderated by a polyethylene layer to produce cold neutrons. The UCN guide containing the solid deuterium converter was contained in the polyethylene cold neutron trap. The UCN guide system led vertically out of the source to a storage bottle above the cryostat that was closed by valves A and B. From valve B, another vertical guide section led down to a helium-3 UCN detector (see below). All the UCN guides were 3.91 cm in radius. Except for the bottle between valves A and B, the guides were stainless steel coated with nickel-58. The bottle section was uncoated stainless steel. The vertical guide section (up to valve A) holding the solid deuterium was 125.3 cm long, containing a volume of 6020 cm³ (including the 240 cm³ of solid deuterium). This length of guide included a 5.1 cm long section at a 45 degree angle to the vertical and a 16.6 cm long horizontal section, so that the bottle was 107.2 cm above the bottom of the guide.  The bottle between valves A and B was 74.7 cm long, for a volume of 3590 cm$^3$.  The guide section leading down to the detector was 160.2 cm long for a volume of 7690 cm$^3$, and included a vertical drop of 130.4 cm between the bottle and the detector.

Measurements of the UCN lifetime in solid deuterium made using this source have been previously reported[18] and found to be in agreement with theory.[19] The longest lifetime measured, 28 ms, was limited by thermal up-scattering, nuclear absorption on residual hydrogen-1, and exothermic para- to ortho-deuterium transitions in the residual para-deuterium in the source. In this letter we report on a significant extension of this previous work to large SD2 volumes and much larger proton pulses in combination. The production rate of UCN is reported and prompt heating of the SD2 by radiation from the spallation target is shown to result in only a small reduction in the rate of UCN production for incident proton pulses up to 95 µC.

Deuterium was prepared in the ortho state using a hydrous iron (III) oxide, $Fe_2O_3 \bullet x(H_2O)$, converter cooled to a few degrees below the triple point at 18.7 K.[20] The deuterium was frozen in the lower part of the cryostat using a liquid helium transfer refrigerator. Both the hydrogen contamination and the para-fraction in the SD2 were measured by means of rotational Raman spectroscopy on a gaseous sample taken by warming the deuterium after the UCN measurements.[20] The para-deuterium fraction for the measurements reported here was determined to be 5.0 ± 0.7% and the hydrogen contamination less than 0.11%. All of the data reported in this paper were taken with an SD2 moderator volume of 240 ± 36 cm$^3$.  The variation of deuterium volume with operating temperature was 0.5%[21], well within the uncertainty quoted above.

The temperature of the solid was monitored using a silicon diode thermometer mounted to the outside of the guide tube. Subsequent measurements, with diodes embedded in the solid, indicate that these measurements were accurate to about ±1 K. For most of the measurements presented here, the SD2 temperature was below 5 K.

However, for the large incident proton pulse runs, the temperatures rose after each pulse. At the start of the 95 μC pulse, the SD2 temperature was 8 K. Based on previous results[18], the UCN lifetime in the solid varies from 17 ms down to 12 ms over this temperature range.

Proton pulses were produced using the Los Alamos Neutron Science Center (LANSCE) 800 MeV linear accelerator. Protons were incident on a tungsten spallation target located 6 cm below the solid deuterium sample, inside the cryostat vacuum. The total charge for each measurement was delivered in one or more pulses, each up to 500 μs long, spread over a 1 second interval. The charge in each pulse was measured by integrating the induced pulse in a current-measuring toroid located 1 m upstream of the SD2 cryostat. The precision of the measurement of the integrated charge in each pulse is estimated to be 5%.

The UCN detector was a wire chamber filled with 1 bar of $CF_4$ and 10 mbar of $^3$He, separated from the UCN guide vacuum by a 0.25 mm thick aluminum window. Neutrons were detected through the $^3$He(n,p)$^3$H exothermic reaction. The low $^3$He pressure reduced the detector sensitivity to thermal neutrons with a smaller reduction in UCN efficiency because of the inverse dependence on neutron velocity of the $^3$He neutron absorption cross section. The $CF_4$ provided a high stopping rate for the charged reaction products so that most of the energy of the reaction was deposited in the detector gas rather than the walls. The detector was operated with a significant gas gain, and the signals were terminated into a 50 ohm resistive load. This highly differentiated mode of operation, with much greater efficiency for detecting slow neutrons than gamma rays, allowed the chamber to recover quickly from the large overload due to prompt radiation following the proton pulse at the expense of a reduction in the neutron detection efficiency by a factor of two. Absorption in the window, up-scatter in the $CF_4$, the solid angle of the active detector region, and electronic inefficiency all contributed to the inefficiency of the detector. These effects were all either measured or simulated using Monte Carlo techniques for our experimental geometry. The total detector efficiency was calculated to be $0.33 \pm 0.03$, averaged over the phase space and velocity spectrum of neutrons arriving at the detector.

A series of measurements was performed to validate the Monte Carlo predictions of the source performance using single pulses of about 0.9 µC of protons. UCN were bottled in contact with the solid deuterium, using valve A shown in Fig. 1. The number of neutrons counted in the detector was measured as a function of the time after the proton pulse at which valve A was opened. The time distributions for one set of proton pulses with different valve opening times, along with the corresponding Monte Carlo predictions, are displayed in Fig. 2, normalized to the number of incident protons per pulse but not corrected for transport or detector efficiency. The summed data as well as Monte Carlo calculations are shown in Fig. 3, in this case including corrections for transport and detector efficiency. The lifetime of UCN in these measurements was dominated by losses in the solid deuterium, as opposed to the neutron lifetime or losses on the walls of the guide system. The UCN lifetime in the solid deuterium used in the Monte Carlo calculations, 12 msec,[18] gives a good account of the data.

The lifetime of neutrons stored in a bottle formed by closing valves A and B was measured in a sequence where valve A (the inlet valve) was initially open and valve B (the outlet valve) was closed. Valve A was closed 1.5 seconds after the proton pulse and neutrons were counted as a function of the time between the closing of Valve A (the beginning of the storage period within our bottle) and the opening of Valve B. These data along with Monte Carlo predictions are also shown Figs. 2 and 3. The normalized but uncorrected arrival time distribution for a single set of runs is shown in Fig. 2; the normalized and corrected integrated data are shown in Fig. 3.

The normalization of the Monte Carlo has been varied to simultaneously fit both the storage bottle and deuterium volume lifetime scans. The quality of the fit leads to a 4% uncertainty in the Monte Carlo estimate of the transport efficiency from the bottle to the detector, which was 0.71. The normalization gives an estimate of the number of UCN produced per incident charge on the spallation target. We estimate 466 ±92 UCN/cm$^3$/(μC of incident protons) for UCN with velocities below 6.6 m/sec in the SD$_2$, the critical velocity at the SD$_2$-$^{58}$Ni interface.  This is reasonably close to a previously reported model prediction of 750 UCN/cm$^3$/μC for the configuration of our apparatus.[22]

We also measured the number of UCN as a function of the size of the proton pulse. For these measurements we increased the proton charge delivered to the target using multiple proton pulses delivered within a 1 second interval. The data were taken with the same valve sequence as the bottle lifetime scan with a storage time of 1.0 s. These data are plotted, as a function of incident proton charge, in Fig. 4. The maximum charge delivered to the spallation target was limited by health physics considerations, as we had no remote handling facilities and radiation shielding in our experimental area was limited.  The UCN density has been calculated by dividing the number of observed neutrons by the volume of the storage bottle, 3590 cm$^3$; by the detector efficiency, 0.33 ± 0.03; and by the transport efficiency, estimated using the Monte Carlo simulation, 0.71 ± 0.03, yielding an additional overall normalization uncertainty of 10%.  The uncertainty caused by the statistics of the fit of the Monte Carlo to the data (Fig. 2) is included in the above uncertainty estimates.

There is a distinct reduction in the ratio of UCN density to incident proton charge at the highest charges.  This is likely to be due to the slow rise in the target temperature through the course of the high current runs. Monte Carlo simulation predicts a 10% decrease in the UCN count with the measured 3 K rise in the starting SD$_2$ temperature and correspondingly shorter lifetime of UCN in the SD2.

The maximum density achieved in the storage bottle, 145 ± 7 UCN/cm$^3$, is significantly larger than peak bottled densities reported previously. A single proton pulse of 95 μC would produce an instantaneous density of 44000 UCN/cm$^3$ in the SD2. Although in principle this is the limiting density that can be extracted from an ideal shuttered source of our geometry using such proton pulses, losses due to absorption on the bottle walls and other causes would likely reduce this considerably. Our model predicts that 1300 UCN/cm$^3$ could be extracted as the limiting density from our source by moving valve A to position C in Fig. 1, so that the volume containing the SD2 is about twice the SD2 volume.   This prediction assumes a 95 μC proton pulse every 10 seconds, with the shutter open for 1 second for each pulse, and with a UCN lifetime of 28 msec in the SD2 (the longest we have measured)[18]. This prediction drops to a density of 889 UCN/cm$^3$ if the wall loss rates are assumed to be 10$^{-4}$/bounce.[13]

These measurements demonstrate a new technology for UCN sources that should allow more than an order of magnitude gain in UCN density over existing UCN sources for practical source bottle geometries.  Optimization of the source geometry, increase in the solid deuterium volume, and increase in the incident proton current are expected to result in even higher produced UCN flux and density in production sources based on the prototype source described here.  Such significant improvements in UCN density would allow a new generation of fundamental measurements with free neutrons to be carried out.

We would like to acknowledge the LANSCE operations staff for delivering beam in the new pulsed modes needed for this experiment and Warren Pierce for his excellent machine-shop support. We also acknowledge fruitful conversations with R. Golub.

This work has been supported by the DOE LDRD program and by the NSF 9420470, 9600202, 9807133, 0071856.

Figure 1. Schematic view of the apparatus used for this experiment. Neutrons were bottled in the region between valves A and B.  SD$_2$ lifetime measurements were made by counting the number of UCN that survived in contact with the deuterium as a function of time using valve A, with valve B left open throughout. Valve C is illustrative of a hypothetical shuttered source.  The figure is not to scale.

Figure 2. Top) Measured counts, normalized to incident proton charge, as a function of time for neutrons stored in contact with the $SD_2$ using valve A. Bottom) Same as the top but for neutrons stored in the bottle using Valves A and B. The solid lines are Monte Carlo calculation normalized to the data as described in the text.  The data are detected counts normalized only to the number of incident protons in each pulse, in 0.16 s time bins.

Figure 3. Top) Corrected total UCN, normalized to incident proton charge, as a function of storage time for neutrons stored in contact with the SD2 using valve A. Crosses are the measured data, circles the Monte Carlo calculation, and the solid line is an exponential fit to the data shown on the plot.  These data are normalized to the number of incident protons in each pulse and also corrected for the estimated transport and detection efficincy.  Bottom) Same as the top but for neutrons stored in the bottle. Error bars show the statistical uncertainty on the data.

Figure 4.  Circles display the bottled density as a function of incident proton charge (left axis). Crosses display the ratio of bottled density to the incident proton charge (right axis).  The red symbols represent a repeated run, performed after the highest incident charge runs, showing only insignificant deviation from the earlier run.  The points in this figure are corrected for transport and detection efficiency.

Correspondence and requests for materials should be addressed to A. Saunders (e-mail: asaunders@lanl.gov).

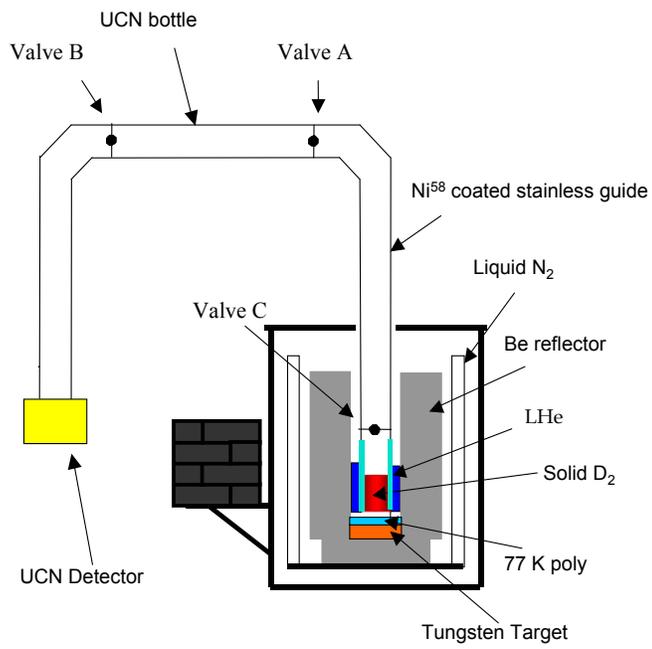

Valve B

UCN bottle

Valve A

Ni$^{58}$ coated stainless guide

Valve C

Liquid N$_2$

Be reflector

LHe

Solid D$_2$

77 K poly

Tungsten Target

UCN Detector

Figure 1.

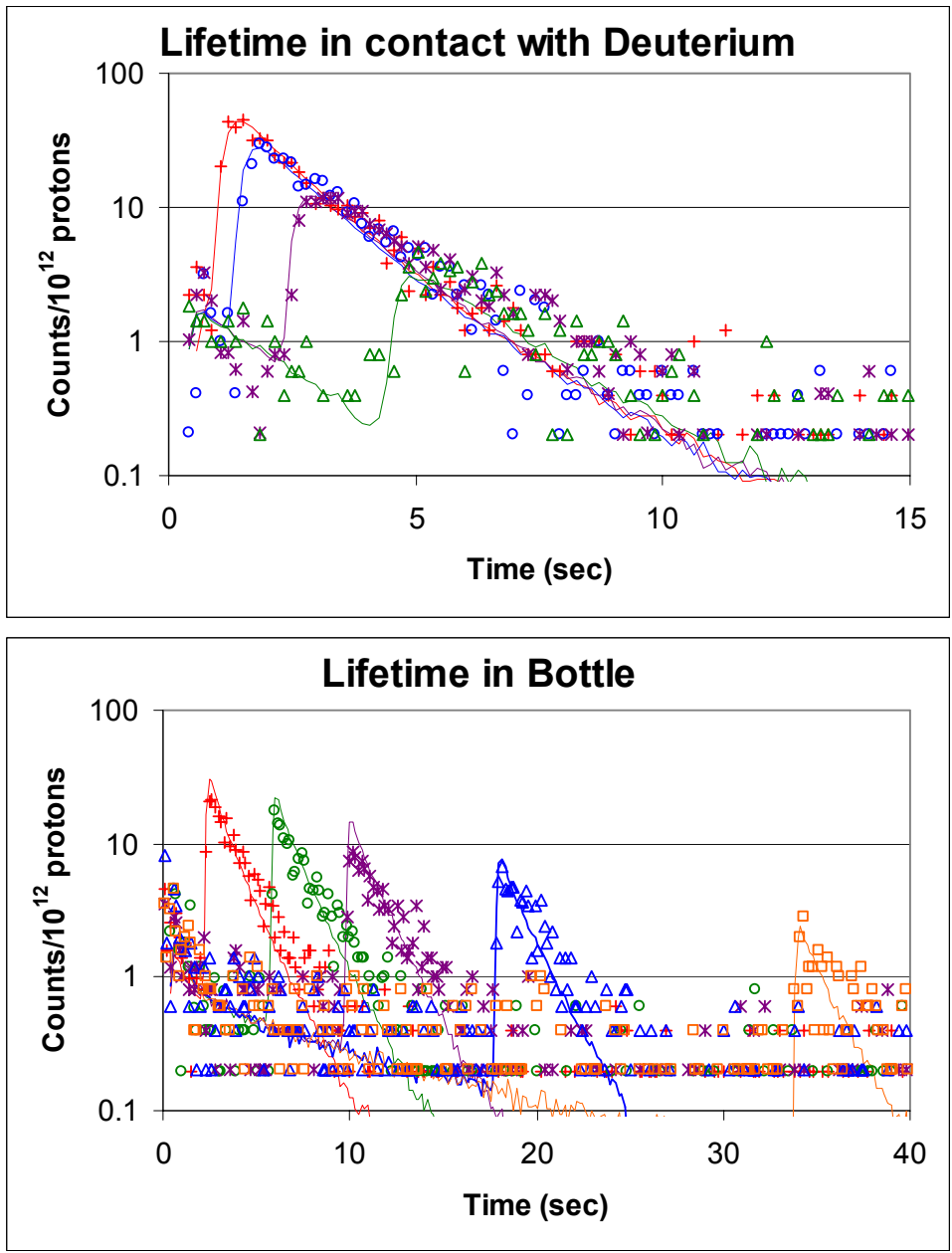

Figure 2.

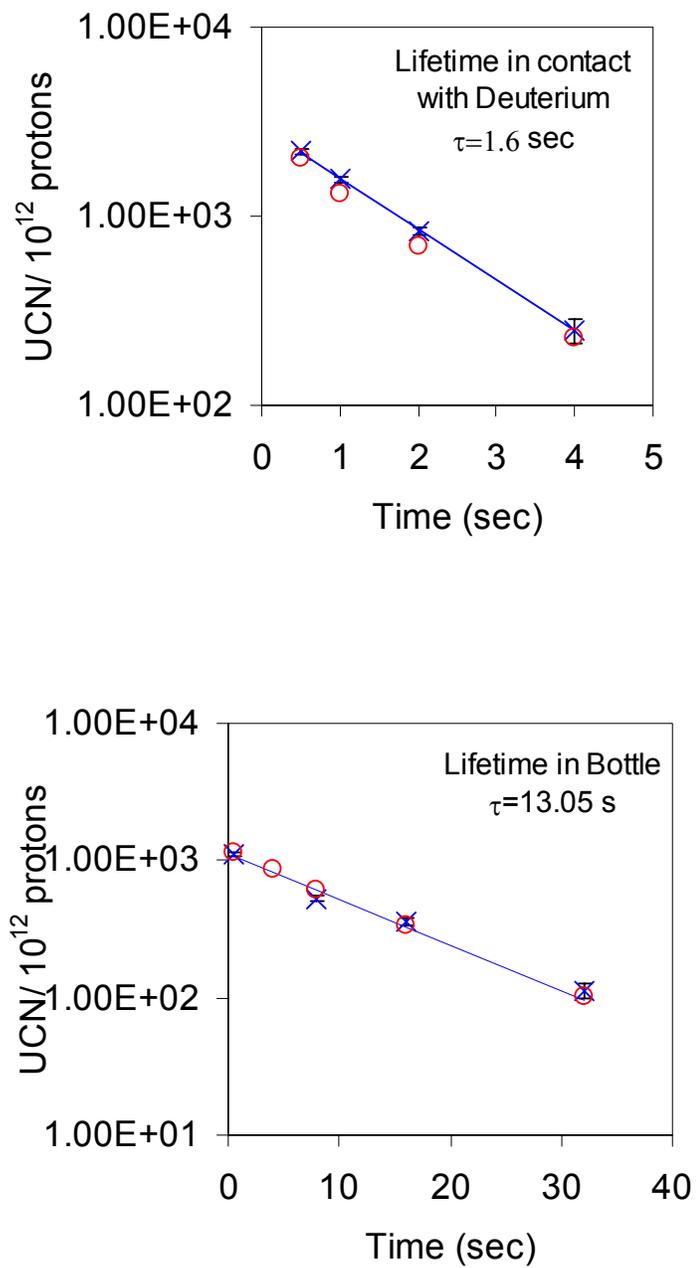

Figure 3.

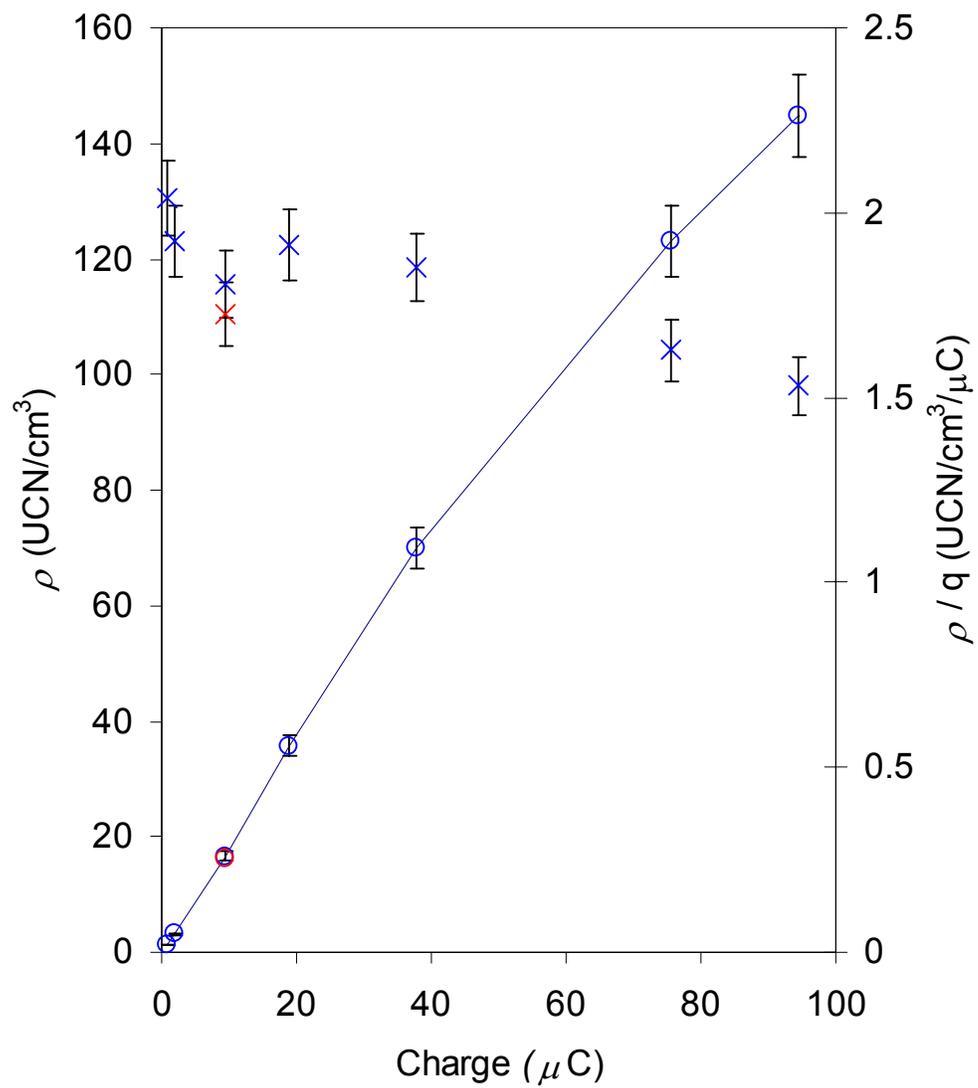

Figure 4.